\RequirePackage{fix-cm}
\documentclass[twocolumn]{svjour3}          
\usepackage{graphicx}

\begin{document}

\title{Topological states in strongly correlated systems
	
}


\author{V. Yu. Irkhin        
	 \and
       Yu. N. Skryabin 
}


\institute{V. Yu. Irkhin \at
              M. N. Mikheev Institute of Metal Physics, 620108 Ekaterinburg, Russia \\
              \email{valentin.irkhin@imp.uran.ru}           
}

\date{Received: date / Accepted: date}

\maketitle

\begin{abstract}
Topological order in strongly correlated systems, including quantum spin liquids, quantum Hall states in lattices and topological superconductivity is treated. Various metallic non-Fermi-liquid states are discussed, including  fractionalized Fermi-liquid (FL$^*$) and phase string theories.
Classification of topological states and differences between quantum and classical topology are considered.


\end{abstract}

\section{Introduction: topological order}
\label{intro}

Since 1980-s, the investigation and classification of a new type of order   in the condensed matter --- of topological order, became a field of intensive researches. 
Main experimental grounds were the discovery of fractional quantum Hall
effect \cite{Tsui}, which is an essentially correlation phenomenon (unlike the integer quantum  Hall effect), and of high-temperature superconductivity in strongly correlated cuprates \cite{Bednorz}. Later, a possible equivalence of both these phenomena was established within resonating-valence-bond (RVB) theory \cite{Kalmeyer}. 

The term ``topological order'' comes historically
from the low-energy effective theory of chiral spin states in topological quantum field theory (TQFT) \cite{Witten} which describes many-body
states with topological degeneracy at low energies. The term ``topological'' here means the long-range entanglement and therefore  refers to quantum topology \cite{Zoo}. It is necessary to distinguish this from the  classical topology,
which treats the vortices in the superfluid liquid,  the difference between the sphere and  torus,  the
vortex in superfluid, etc. \cite{Mermin}. 
Quantum fluctuations eliminate the degeneracy of the ground state of strongly frustrated systems by quantum tunneling, leading to non-degenerate ground states (up to global symmetry transformations or topological degeneracies). 
One  distinguishes topological order (elementary excitations with an energy gap) and quantum order (the more general case of gapless excitations). 




The emergence of correlations in topological systems is rather uncommon. Here, there are no usual ``energetic'' correlations (precursor of long-range order), but specific correlations appear, which are caused by the topology of the sample. In this case, the ground state is degenerate owing to the topological characteristics and not because of symmetry. The simplest example: the system feels
the influence of a single pricked point (the transformation of a sphere into a torus).

From the macroscopic viewpoint, the topological
order is characterized by the strong ground-state
degeneracy. 
Topological degeneracy can be realized in protected	two-level states -- qubits, which enables one to perform topological quantum computation \cite{Zeng}.
On the other hand, from a microscopic viewpoint the topological order is  the state of matter with gapped energy spectrum \cite{1406.5090}, which is not reduced to the product of one-particle eigenstates, if we do not take into account phase transitions with closing  the energy gap. This means long-range correlations, i.e., entanglement. 

Topological phases are phases of matter at zero temperature. 
They are disordered liquids that seem to have no simple characterization, but  actually can have rich patterns of many-body entanglement representing new kinds of order. 

Authors of Ref. \cite{cre,cre1} propose the following classification of gapped topological entangled  phases. For gapped quantum systems without any
symmetry, one has two classes: short-range entangled (SRE) states and long-range entangled (LRE) states. SRE states  can be transformed into direct product states via local unitary transformations (LUT). All SRE
states can be transformed into each other via LUT,
and therefore belong to the same phase.
On the other hand, LRE states cannot be transformed into
direct product states via LUT. 
A number of LRE states also cannot be transformed
into each other, so that they belong to different classes
and represent different quantum phases. These 
quantum phases are just the topologically ordered
phases, including Fractional quantum Hall states, chiral
and quantum spin liquids. 
Such different topological orders can result in a quasiparticle spectrum 
with fractional statistics and fractional charges.

For gapped quantum systems with symmetry, the structure of phase diagrams is much richer. The corresponding LRE phases are called Symmetry Enriched Topological (SET) phases and are described by projective symmetry
group (PSG).  Symmetry breaking (SB) and LRE can appear together forming  SB-SRE states.  An example is given by the topological superconducting states.
Long-range-entangled states that do not break any symmetry can also belong to different symmetric phases.
Examples are Z$_2$ symmetric spin liquids with spin rotation, translation, and time-reversal symmetries.
 
Not only LRE, but also SRE states  with symmetry can belong to different phases.  Here, there are two possibilities:

(i) the usual Landau SB states which belong to different
equivalent classes of LUT and have different broken symmetries;

(ii) states with short-range entanglement can belong to different equivalent classes of the symmetric LUT even if they do not break any symmetry of the system, i.e. have the same symmetry. Such phases are beyond Landau symmetry breaking theory and are called symmetry protected topological (SPT)
phases. An example is given by the band and topological insulators which have the same symmetry, but belong to two different equivalent classes of symmetric LUT.




In the present paper we review various topological states in strongly correlated systems including quantum spin liquids,  quantum Hall states and topological superconductivity. Some further details can be found in the reviews \cite{Scr2,Scr3}.

\section{Quantum spin liquids}
\label{sec:1}

Quantum spin liquids (QSLs) are phases
of matter, which are characterized by emergent dynamic gauge fields and topological order, and fractional excitations
interacting with the gauge field. The existence of nontrivial many-particle states is confirmed by the Lieb--Schultz--Mattis theorem \cite{Lieb} and its higher-dimensional generalization by Hastings \cite{Hastings}. This states that in the system with the half-integer spin per cell and global U(1) symmetry the excitation spectrum in the thermodynamic limit cannot satisfy simultaneously two requirements: (a)  the ground state is unique; and (b)  there is a finite gap for all excitations. This means that the  state with the gap and unbroken symmetry should have a degeneracy of a topological  nature.

According to  \cite{wen2002quantum}, spin liquids
can be divided into four classes: (i) Rigid spin liquid where spinons (and all other excitations) are fully gapped
and may have Bose, Fermi, or fractional statistics; (ii) Fermi spin liquid where spinons are gapless and
are described by a Fermi liquid theory. (iii) Algebraic spin liquid where the spectrum is gapless, but excitations are  not
described as free quasiparticles; (iv) Bose spin liquid where low-lying gapless excitations are described by a free-boson theory. 

Being examples of quantum-ordered states, different spin liquids cannot be distinguished by their symmetry properties. The first discovered topological invariants  that define topological order were (i) the robust ground state
degeneracy on a torus and other closed space manifolds, (ii) the non-abelian geometric phases  of the degenerate ground states, and (iii) the chiral central charge $c$ of the edge states. The ground states of gapped spin liquids always have topological degeneracy, which cannot be related to any symmetry. Some  stable quantum phases have gapless excitations even without any spontaneous symmetry breaking. In particular, such excitations in algebraic spin liquids interact in the limit of zero energy, but the interaction does not open an energy gap. Thus the quantum order (and not symmetry) protects the gapless excitations and makes algebraic spin liquids and Fermi spin liquids stable \cite{Wen22,Zoo}. 


\begin{figure}[h]
	\includegraphics[width=0.45\textwidth]{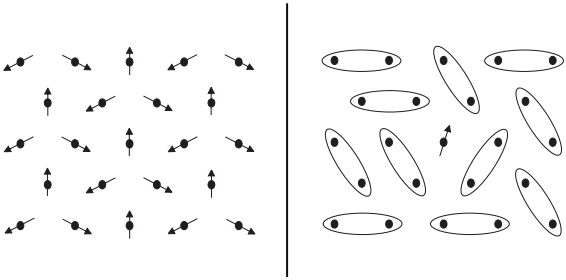}
	\caption{Quantum phase transition on a frustrated triangular lattice
		according to Ref.\cite{Sachdev2}.
		The state on the left has 	120$^{\circ}$	non-collinear magnetic order, 
		The state on the right is a ``resonating valence bond'' (RVB)
		paramagnet with topological order and fractionalized $S=1/2$
		neutral spinon excitations,  one spinon being shown in the centre. In the ground RVB state, each site participates in one bond. When a bond is broken, two unpaired sites appear, which possess 1/2 spins, the corresponding excitations (spinons) being not charged. At the same time, the vacant site (hole)  carries a charge, but not a spin.}
	\label{2}
\end{figure}

The spin liquid can demonstrate some strange properties. 
Sometimes excitations can  carry fractional statistics.  
Generally, topological systems can have exotic spectrum of quasiparticle  excitations (neutral spin-1/2 fermions -- spinons, charged bosons -- holons, see an example in Fig.1). 
A description of gapless quantum spin liquids in strongly correlated metals is provided by various  theories using slave (auxiliary) representations for such quasiparticles \cite{Nagaosa,Weng,Sachdev}.

The exotic quasiparticles  were first introduced by Anderson \cite{633a} in the theory of two-dimensional cuprates as
\begin{equation}
	\tilde{c}_{i\sigma }=X_i(0,\sigma )=b_i^{\dagger }f_{i\sigma }.
	\label{eq:6.131c}
\end{equation}
According to Anderson, spinons  $f_{i\sigma}$ are fermions  and holons $b_i$ are bosons.  This choice of statistics is not unique and can be changed depending on the physical problem, e.g., on the presence or absence of antiferromagnetic ordering, see Ref.\cite{Punk1}.
The representation (\ref{eq:6.131c}) leads to difficulties in simple mean-field treatments, since the problem of non-physical states occurs: the on-site no-double-occupancy condition is violated.
Thus spinons and holons become coupled by a gauge field
required to satisfy this constraint.

A more general SU(2) representation \cite{Nagaosa} exploits two types of bosons,
\begin{eqnarray}
	\tilde{c}_{i\uparrow }& ={\frac{1}{\sqrt{2}}}\left( b_{i1}^{\dagger
	}f_{i\uparrow }+b_{i2}^{\dagger }f_{i\downarrow }^{\dagger }\right),  \nonumber
	\\
	\tilde{c}_{i\downarrow }& ={\frac{1}{\sqrt{2}}}\left( b_{i1}^{\dagger
	}f_{i\downarrow }-b_{i2}^{\dagger }f_{i\uparrow }^{\dagger }\right).
	\label{22}
\end{eqnarray}

Upon the treatment of strongly correlated compounds, e.g., of copper–oxygen high-temperature superconductors, the $t-J$ model (the Hubbard model with the on-site Coulomb repulsion $U\rightarrow \infty $ and including the Heisenberg exchange) is widely used. The Hamiltonian of this model in the many-electron representation of Hubbard's operators $$	X(\Gamma ,\Gamma ^{\prime })=|\Gamma \rangle \langle \Gamma	^{\prime }|$$ (where $\Gamma =0, \sigma$)
takes  the following form:
\begin{equation}
	\mathcal{H}=-\sum_{ij\sigma }t_{ij}X_i(0,\sigma )X_j(\sigma,0)
	+\sum_{ij}J_{ij}{\bf S}_i {\bf S}_j.
	\label{eq:I.7}
\end{equation}
In the representation of auxiliary bosons   $ X_i(\sigma 0)= f_{ i\sigma}^\dagger b_{ i}$, one can  introduce the pairings
\begin{equation}
	\chi_{ij} = \sum_\sigma \langle
	f^\dagger_{i\sigma} f_{j\sigma} \rangle, \,\,
	\Delta_{ij} = \langle  f_{i\uparrow}f_{j\downarrow} - f_{i\downarrow}f_{i\uparrow}  \rangle .
	\label{Eq.40}
\end{equation}

Anderson et al \cite{633a,Baskaran}  used the slave-boson approach to construct a uniform RVB state -- a symmetric spin liquid with all the lattice symmetries  (SU(2)-gapless state). 
In the uniform RVB phase (uRVB), $\chi_{ij} = \chi$ for all bonds and is real, and the gap $\Delta_{ij}$ is zero, so that the spectrum of $f$-fermions has the form $E_{\bf k} = - 2 {J}\chi (\cos k_x + \cos k_y)$.

Later two more spin-liquid states were constructed using the same U(1) slave-boson approach (for details, see \cite{Nagaosa}). One is the $\pi$-flux ($\pi$fL) phase which is a SU(2)-linear symmetric spin liquid with the spectrum
\begin{equation}
	E_{\bf k}=\pm \frac34 J|\chi| \sqrt{\sin^2 k_x + \sin^2 k_y}.
\end{equation}
The other is staggered-flux/d-wave state (staggered flux liquid, sfL) which is a U(1)-linear symmetric state. This phase demonstrates a linear Dirac  spectrum at the points $(\pm \pi/2,\pm \pi/2)$, 
\begin{equation}
	E_{\bf k}=\pm \frac34 J\sqrt{\chi^2 (\cos k_x+\cos k_y)^2
		+ \Delta^2 (\cos k_x - \cos k_y)^2 }.
\end{equation}

Generally, the U(1) and SU(2) spin liquids are  unstable at low energies and cannot be treated as the ground states. However, they can
provide a good description in a broad intermediate region of length and time scales close to a quantum phase transitions.
Thus the first known stable spin liquid is the {\it chiral spin liquid} where the spinons and holons carry
fractional statistics and true spin-charge separation occurs. Such a state violates the time reversal and parity symmetries and is a SU(2)-gapped state.
The coupling between the slave fermions and the gauge field in the chiral state  is identical to the coupling between the electrons and the electromagnetic field. Therefore the system  
has properties similar to the Hall effect \cite{Wen22}.
 The SU(2) gauge fluctuations in the chiral spin state do not result
in an instability since the gauge fluctuations are suppressed and become massive due to the Chern-Simons term \cite{Wen02}.

A way to obtain a stable deconfined phase is lowering the U(1) or
SU(2) symmetry down to a  Z$_2$ gauge structure \cite{Nagaosa}.
Such a phase is called a Z$_2$ spin liquid or a short-range RVB state.
The corresponding low-energy effective theory deals with massless Dirac fermions and fermions with small  Fermi surfaces, coupled to a Z$_2$ gauge field. Since the Z$_2$ gauge interaction is irrelevant at
low energies, the spinons are free fermions at low energies and we have a true spin-charge separation in the Z$_2$-gapless spin liquid.

The state with noncollinear SU(2) flux is a Z$_2$ state where all the gauge fluctuations possess a gap. Here the fluctuations mediate \textit{short-range} interactions between fermions and do not radically change the properties of the mean-field solution, the gauge interactions being irrelevant. 
This implies  the existence of a true physical spin liquid which contains  fractionalized spin-1/2 neutral Fermi excitations -- spinons. Such a spin liquid also has Z$_2$ vortex excitations, so-called visons. The bound state of a spinon and a Z$_2$ vortex gives us a spin-1/2 Bose excitation.
The ground state degeneracy of the Z$_2$  RVB state is a reflection of its long-range entanglement. Although all spin-spin correlation functions decay exponentially  in the ground state, there are  Einstein–Podolsky–Rosen-type long-range
correlations, so that the quantum state retains information  about the global topology  \cite{Sachdev12}.



Fluctuations of the gauge field are physically chirality fluctuations or fluctuations of orbital current. The corresponding staggered flux  (SF) phase is obtained in the slave-boson mean-field approach \cite{Nagaosa}. The SF state is competing with $d$-wave superconductivity and antiferromagnetic (AFM) ordering in systems with nodal points (high-$T_c$ cuprates).
The SF state  provides an example of collinear SU(2) flux state which is invariant only under a U(1) rotation. This is a marginal situation.




Another way to get a deconfined phase is to make the gauge boson massive.
As mentioned above, most simple example of such a situation is provided by  the chiral spin liquid. 
The picture is as follows  \cite{Wen22}.
The  excitation in the mean field approximation is obtained by inserting a spinon into the conduction band. However, this excitation itself is not  physical, since this insertion violates the constraint $\sum_{\sigma} \langle f^\dag_{i\sigma}f_{i\sigma}\rangle=1$.
The additional density of spinons can be eliminated by adding a vortex flow of the gauge field
\begin{equation}
	\Phi= -\pi \sum_{i} \left(\sum_{\sigma}\langle f^\dag_{i\sigma}f^{}_{i\sigma}\rangle-1\right).
	\label{fi}
\end{equation}
Thus the physical quasiparticles are spinons dressed by a $ \pi $ vortex which carry spin 1/2. At the same time, spinon, which bears a unit charge of the gauge field, has a fractional (semion) statistics, being a bound  state of charge and vortex \cite{Wen22}. 

\begin{figure}[h]
	\includegraphics[width=0.45\textwidth]{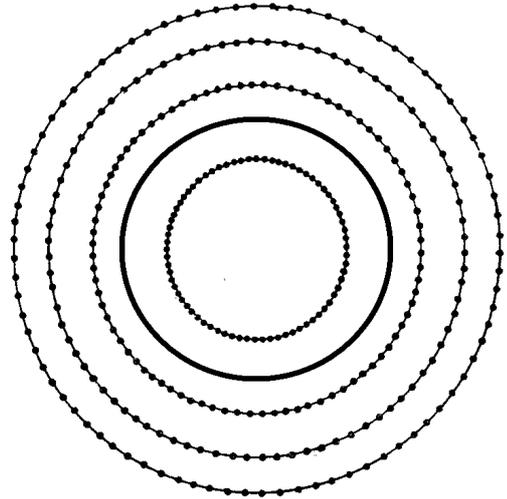}
	\caption{Landau quantization in the gauge field in the $k_x-k_y$ plane. The points indicate density of electron states on the levels. The bold line is the Fermi energy in the insulating state.}
	\label{25}
\end{figure}

In the presence of fluxes, quantization in the gauge field gives rise to Landau-type levels  \cite{Wen22,Scr4} (see Fig. \ref{25}). In this case, the zero level has a degeneracy equal to the number of the flux quanta. The addition of a flux quantum yields a zero fermionic mode for each type of Dirac fermions. After inclusion the crystal potential, the Landau levels are transformed into narrow correlated bands. In this sense, the Hubbard bands (which can be described in the simplest Hubbard-I approximation, including for degenerate bands  \cite{Hubbard,Hubbard2}, as broadened atomic levels) are spinon bands. It can be hypothesized that the Hubbard broadening of levels (for example, due to scattering and resonant broadening \cite{Hubbard3})
plays a role similar to that of disorder which is crucial for the quantum Hall effect. 

A similar theory was developed by Levin and Wen \cite{Levin}  within the doubled Chern-Simons theories
where macroscopic chirality is absent and time reversal is not broken.
They treated the  mutual statistics of spinon and vison excitations which can be considered as mutual semions (see also recent work \cite{Levin1}).
Here,  encircling a spinon around a vison  yields a Berry phase of $ \pi$,  the charges moving on the square lattice, while the fluxes  on the dual lattice. Some corresponding three-dimensional models can be also  formulated \cite{Levin}.

In Weng's phase string theory  \cite{Weng},
mutual semion statistics of spinons and holons in superconducting and AFM phases occurs within slave-fermion approach in the $t-J$ model.
In the underdoped regime the AFM and superconducting phases are dual: in the former, holons are confined while spinons are deconfined,
and in the latter vice versa. 
The  mechanism of non-Fermi-liquid-like behavior is owing to Anderson's  unrenormalizable Fermi-surface phase shift produced by the doped holes, which is related to a many-body Berry-like phase \cite{Weng1}.
Weng's Lagrangian possesses both parity and time-reversal symmetry. Note that this structure is similar to that for Z$_2$ spin liquid including spinons and visons \cite{Sachdev2}.

For an arbitrary loop $C$ on the lattice  one has \cite{Weng}
\begin{eqnarray}
	\sum_{\langle i,j \rangle\in C}\, A_{ij}^s &  =  & \pi \sum_{l\in
		S_C}\, \left( b^\dagger_{l\uparrow}
	b^{}_{l\uparrow}-b^\dagger_{l\downarrow} b^{}_{l\downarrow}\right) \,\, {\rm mod}\, 2\pi\,,\nonumber\\
	\sum_{\langle i,j \rangle\in C}\, A_{ij}^h &  =  & \pi \sum_{l\in
		S_C}\, h^\dagger_l h^{}_l \,\, {\rm mod}\, 2\pi \,, \label{66}
\end{eqnarray}
where $b$ are Bose (Schwinger) spinons, Bose holons $h$ are obtain by bosonization, $A_{ij}^{s,h}$ are the gauge fields, the sum on the left-hand side runs over all the links on
$C$, and on the right-hand side over all the sites
inside $C$. Thus the holon (spinon) bears a $\pi$-flux and couples to the motion of spinons (holons) via  the field $A_{ij}^h$ ($A_{ij}^s$).
The quantization of the flux of the gauge fields 
 results in integer values of spinon and holon occupation numbers.
The mutual Chern-Simons term in \cite{Weng} enables one to realize the no-double-occupancy condition, i.e. to treat properly Hubbard's projection.

Since in the presence of Hubbard projection the Fermi statistics is not applicable, Weng proposed ``the phase string statistics''  describing the  doping picture in the $t-J$ model \cite{Weng2007,Zaanen-Overbosch}. In the undoped case (for the Mott insulator) the fermionic sign structure on non-frustrated lattices  is completely reducible (removable), since the problem of interacting electrons is simply the localized-spin problem with distinguishable electrons (however, for a frustrated lattice, this problem can also include the problem of sign). 
A doped hole remains stable because of the confinement of spinons and holons by the field of the phase shift, in spite of the fact that the background is a spinon--holon sea. The exact deconfinement appears only in the limit of the zero doping, when the hole is decomposed into a holon and a spinon. At doping, there appears an irreducible sign structure, but for low doping this structure is very rarefied in comparison with the equivalent system of free fermions. 

A way to treat the Weng statistics is systematically calculating the irreducible signs in terms of world lines \cite{Zaanen-Overbosch}. The irreducible signs entering into the partition function of the $t-J$ model can be represented as
\begin{equation}
	Z_{t-J} = \sum_c (-1)^{N^h_{\rm ex}  [c] + N^{\downarrow}_h [c]} {\cal Z} [c].
	\label{tJsigns}
\end{equation}
The sum  is taken over the configurations of the world lines $c$ of spinons and holes. Relative to each other, the holes behave as fermions,  and $N^h_{\rm ex} [c]$ numerates the exchanges in the configuration $c$. The spinons  are bosons with a solid core, the up spins being considered as the background. The sign structure is determined by the parity of the number of hole--spinon collisions for a concrete configuration of world lines.

The projected annihilation electron operator  is written in the slave-fermion representation, the sign coefficient $(-\sigma)^j$ being included explicitly:
$$ {\tilde c}_ {j \sigma} = (-\sigma)^jf^{\dagger}_j b_{j\sigma},$$ where $f^{\dagger}_j$ is the Fermi holon, and the bosons $b^{\dagger}_{j\sigma}$ are used for the spin system. Because of the sign coefficient, the spin-spin superexchange acquires the general negative sign. Therefore, the  ground state wave function  for the purely spin system in the half filled case does not contain nodes.

Fractionalized Fermi liquid (FL$^*$) picture \cite{Sachdev,Sachdev1} presents another scenario for anomalous frustrated metallic systems. This was initially formulated within the framework of the $s-d(f)$ exchange (Kondo lattice) model. 
This theory starts from a fractionalized spin liquid with exotic excitations and adds
conventional carriers in a second band.  Low-temperature properties are Fermi-liquid-like or not, depending on that the  excitations of the spin-liquid subsystem are gapped or gapless. 
FL$^*$ state  may demonstrate a number of instabilities at lowering temperature, 
including AFM ordering and unconventional superconductivity.
In this state, which is a  metallic spin liquid of topological nature, charged excitations have conventional quantum numbers (charge $\pm e$ and spin 1/2), but  coexist with additional fractionalized degrees of freedom in the second band. The  FL$^*$ concept was applied to frustrated Kondo lattices \cite{Vojta}. 
The most important topological feature of FL$^*$ is that violates  the Luttinger theorem -- has the small Fermi surface (FS) which does not include localized $d(f)$-electrons, unlike the  ground state of the Kondo lattice -- heavy Fermi liquid (FL). In this state, fermionic spinon excitations  form a ``ghost'' Fermi surface.

In the FL$^*$ phase, the anomalous Kondo pairing, i.e., the condensation of Higgs boson $b$ is absent, $\langle b_i \rangle=\langle f^\dagger_i c_i\rangle=0$. At the same time, there is an anomalous RVB-type average at different lattice sites,  $\chi_{ij}=\langle f^\dagger_i f_j\rangle$. Thus the localized spins do not take part in the formation of the  Fermi surface  (however, they form a separate spinon FS). Owing to the heritage of  the spin liquid,  FL$^*$ state possesses  deconfinement exotic excitations. In the space dimensions $d \geq 2$, a Z$_2$ spin liquid type is stable, and at $d \geq 3$ a U(1) spin liquid can  exist. In this phase,  electronic specific heat is not linear, but  $C/T$ diverges logarithmically \cite{Sachdev}. 
Against the background of this state, there can occur a magnetic instability for the spinon, rather than electron FS, which leads to an itinerant-magnetic state with the spin-density wave, SDW$^*$. 

As follows from the Lieb-Schulz-Mattis theorem, a  state with a gap and unbroken symmetry should have a ground-state degeneracy which has a topological nature. In the two-dimensional case,  the existence of topologically different sectors. being supported by existence of gapped vison excitations on a torus, protects the FL$^*$ state.
Usually the FL$^*$ theory is applied to 2D case, but it can be generalized to 3D case due to Oshikawa’s topological analysis \cite{Oshikawa,Sachdev}.
The FL$^*$ state was treated for near  both  the metal-insulator transition  and  transition between metallic states with large and small Fermi surfaces.   The corresponding phase diagrams were built in both doping case and the situation of interaction-driven transition \cite{Sachdev,Senthil1,Senthil2}.

In the FL$^*$ state, there are  low-energy excitations of local moments, which give a  topological contribution to a change in the momentum of the crystal. Indeed, the action of a vortex flux is analogous to the Lieb--Schultz--Mattis transformation \cite{Lieb}, so that the spin-liquid state in the dimensionality $d = 2$ acquires this momentum change. 
According to the treatment \cite{Oshikawa}, based on the threading the flux quantum  
and on the global gauge symmetry U(1) (charge conservation), 
 the existence of a nonmagnetic FL$^*$ state with small FS is permitted if we include  global topological excitations which naturally appear in the gauge theories. Therefore, the violation of the Luttinger theorem must be accompanied by the appearance of a topological order \cite{Sachdev}. Generally, the formation of a small FS, of the Hubbard splitting, and of a state with disordered local moments can be connected with the topological order in a spin-liquid-like state.
The Oshikawa theorem \cite{Oshikawa} holds for both the metal and insulator, so that we can deal with the Mott-Hubbard state too.






An exotic type of quantum order occurs in the string-net condensation theory \cite{Levin,Zoo}. Such string states  are
similar to Bose-condensed ones, but the condensate
is formed here from extended objects. 
The collective excitations in this condensate 
are not usual Bose particles: the vibrations of closed strings generate gauge bosons. Moreover, at breaking of strings, their ends  give fermions. The quantum topology describing such internal degrees of freedom can be reformulated as a field-theory of closed strings, which are constructed from local spins or  pseudospins (qubits \cite{Zeng}). In turn, one can to pass  from these strings to the concept of ``electric'' and ``magnetic'' gauge fields \cite{Kogut}.

\section{Quantum Hall states} 



Similar to a quantum chiral  spin liquid,
in the quantum Hall effect (QHE) situation  we deal with a gapped essentially topological matter including topological order. Owing to occurrence of a gap and topological degeneracy,  in the effective magnetic (e.g., gauge) field  $T$-symmetry  becomes violated, at least dynamically, and the Chern–Simons term arises which represents charge-flux attachment. 
It is important that any system with nonzero Chern number includes gapless edge modes. These modes, being initially discovered in the integer quantum Hall effect,  are chiral, i.e., propagate only in one direction \cite{Kitaev}.

Low-energy effective theories of fractional quantum Hall (FQH) states are U(1) Chern–Simons theories which describe universal properties.
The dependence of the ground-state degeneracy on the topology of the space indicates the existence of a long-range order in FQH liquids, despite the absence of long-range correlations for local physical
operators. Thus FQH liquids are characterized by hidden long-range orders \cite{Wen02}.

In the QHE situation, electrons in a magnetic field move  along circular cyclotron orbits, and also around other electrons. Then the Landau level number is determined by the number of wavelengths on a given circle. The description of strongly correlated systems requires the formation of flat-band structure, which is similar to system of the  Landau levels. 
Band structures with a gap can be classified topologically
by considering the equivalence classes of the Bloch Hamiltonian that can be continuously deformed into one
another without gap closing \cite{Hasan}.
 In this case, the total Berry flux over the Brillouin zone is (the sum of the integrals of the Berry curvature over all occupied bands) is a Chern topological invariant $C$ which is the
total Berry flux in the Brillouin zone. 
Thus we obtain the Chern insulator which has gapless edge states, similar to topological insulators (however, im the latter case the topological case is invertible -- there are no topological excitations \cite{Zoo}). 
These states can be treated as a consequence of the cyclotron orbit bounce off the edge. The electron states responsible for this motion are chiral.

QHE in two-dimensional (2D)  systems is usually related to the presence of an external  magnetic field, which splits the electron spectrum  into Landau levels. However, QHE may also result from breaking  time-reversal symmetry (i.e., by a magnetic order) without a net magnetic flux through the unit cell of a periodic 2D system 
\cite{Haldane}.

There is another  topological class of insulating band structure  where, although the $T$ symmetry is not broken initially,  the  spin--orbit interaction is present \cite{Hasan}. Such a 2D topological insulator is called a quantum spin Hall insulator. This system is described by the so-called double Haldane model \cite{Haldane} where the  Hall conductivity has opposite signs for up and down spins. In an applied electric field, the up and down spins give Hall currents that flow in opposite directions. Thus, the total Hall conductivity is zero, but there occurs a spin current and a  spin Hall conductivity which is quantized. Taking into account the complex amplitudes of hopping between next-nearest neighbors, a gap opens at the Dirac points, so that the $T$ symmetry becomes broken. Thus we have two bands with Chern numbers (of $C= \pm 1$) and, in the case of half filling, the integer QHE. At some partial fillings, a the fractional QHE  can arise, i.e. topological state of a quantum incompressible liquid  \cite{Neupert}. 

As first stated in Ref.\cite{Tang}, appropriate combination of geometric frustration, ferromagnetism, and spin-orbit interactions can give rise to nearly flat bands with a large band gap and nonzero Chern number.
Such  bands (which mimic Landau levels) can give rise to fractional QHE even at high temperatures. 

The realization of topological Hall phases  on a lattice  in the absence of an external magnetic field  (the ``anomalous Hall effect'') requires a number of simultaneous conditions. The first one is the presence of a nearly  dispersionless (``flat'') bare energy band with nontrivial topology (non-zero Chern number), which provides the picture similar to that of the Landau level (the quasiparticle kinetic energy  is quenched in such a topological band). The second condition is a strong interelectron interaction violating the Fermi liquid picture. Strong correlations are especially important for the fractional Hall effect, when the Landau levels are highly degenerate. 
Since the Hall conductivity is odd with respect to time reversal, topologically nontrivial states can arise when the  $T$ symmetry becomes violated. 

In a number of papers, attempts were made to take into account strong interactions in a  frustrated system in order to obtain phases with topological order within the framework of simple  mean-field-type approximations. In particular, the anomalous QHE can occur dynamically  in the generalized Hubbard model on a honeycomb lattice and other  systems with a quadratic zone intersection point, including the  diamond and kagome lattices. However, later  numerical studies did not confirm the occurrence of exotic topological phases predicted by mean field theories. The matter is that, instead of causing spontaneous $T$-symmetry breaking, strong interactions also tend to stabilize competing long-range orders breaking translational symmetry (see the discussion in \cite{Zhu}).
Thus, when describing lattice QHE systems, it would be correct to start immediately from a strongly correlated state.

\begin{figure}[h]
	\includegraphics[width=0.45\textwidth]{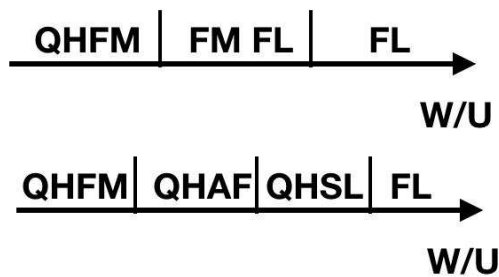}
	\caption{Two possible phase diagrams for spinful Chern band at integer filling $\nu_T = 1$ according to \cite{Zhang}.
		QHFM, QHAF and QHSL stand for quantum Hall ferromagnetism, quantum Hall antiferromagnetism and quantum
		Hall spin liquid. QHSL passes into QHAF through a continuous transition. In the	QHAF state, partial spin polarization coexists with antiferromagnetic order.}
	\label{222}
\end{figure}

In the presence of interaction, the system can be described within the framework of the Hubbard model with the band width $W$ and the Coulomb repulsion $U$. In the case of the Chern zone, the charge cannot be localized (``the obstruction of the Wannier states''), so the narrow band regime cannot be treated in the pure spin model \cite{Zhang}. 
For small $W/U$ ratios, the charge gap is determined by $U$ and is not necessarily removed by the spin disorder. 
A  destruction of quantum Hall ferromagnetism by the kinetic term is possible at integer filling of the Chern band. For intermediate $W/U$, the charge subsystem is still a Chern insulator with quantized Hall conductivity, but the spins are not ferromagnetically ordered. Then quantum Hall antiferromagnetic and quantum Hall spin-liquid phases occur where antiferromagnetic order or a quantum spin liquid coexist with quantized Hall conductivity. Therefore, a transition from a ferromagnetic state with the QHE to a Fermi liquid through unusual antiferromagnetic and spin-liquid phases is possible. The corresponding phase transitions are treated in Ref.  \cite{Zhang}, see Fig.\ref{222}.

The effective Lagrangian describing  QHE for electrons in a magnetic field with inclusion of the Chern--Simons term has the  form \cite{Wen2,Wen22}: 
\begin{equation}
	{\cal L}_{CS} = -
	\frac m {4\pi} \epsilon^{\mu\nu\lambda}a_{\mu}\partial_\nu a_{\lambda}
	- \frac e {2\pi} \epsilon^{\mu\nu\lambda} A_{\mu}\partial_\nu a_{\lambda}.
	\label{eq:C2}
\end{equation} 
Here ${ a}_\mu$ is the internal gauge field, ${ A}_\mu$ is the vector potential of the external electromagnetic field, and $\epsilon $ is the  antisymmetric second-rank tensor. The filling is given by $\nu=1/m$,  $m$ being the charge of the gauge field, i.e., the number of wavelengths in the situation where one electron encircles another.

Equation (\ref{eq:C2})  describes only the linear response of the ground state to the external  field. To obtain a more complete description of a  fractional quantum Hall liquid, one has to introduce excitations. Although in the Laughlin fractional Hall ground state the electron is a fermion, the excited states of the system have a Bose type. Then $m$ is an even integer for Bose states and  odd for Fermi states. A fractional quantum Hall liquid contains two types of quasiparticles: a quasihole (or vortex) in the original electron condensate and a quasihole (or vortex) in the new Bose condensate. 
Introducing the gauge field ${\tilde a}_\mu$ which describes a bosonic current we can represent the total Lagrangian in a  matrix form which is analogous to (\ref{eq:C2}): 
\begin{equation}
	{\cal L} = -
	\frac 1 {4\pi} \epsilon^{\mu\nu\lambda}K_{IJ}a_{I\mu}\partial_\nu a_{J\lambda}
	- \frac e {2\pi} \epsilon^{\mu\nu\lambda}q_{I}A_{\mu}\partial_\nu a_{I\lambda}.
	\label{eq:C3}
\end{equation} 
Here  $(a_{1\mu},a_{2\mu})=(a_{\mu},{\tilde a}_\mu)$, and the matrix $K$ reads
\begin{equation}
	K=\left(
	\begin{array}{cc}
		p_1 &-1\\
		-1&p_2
	\end{array}
	\right)
	\label{eq:C_2_K}
\end{equation}
where $p_1$ is the starting number  $m$ for fermion electronic states, $p_2$ is an even number describing the bosonic field, $\bf q$ is the charge vector, and ${\bf q}^T = (q_1,q_2) = (1,0)$. The occupation numbers now are $\nu = {\bf q}^T K^{-1}{\bf q}$. 
This construction corresponds to the doubled Chern-Simons theory for the topological interaction among both quasiparticles and  vortices \cite{Oganesyan}.

The topological theory of abelian phases of a general gapped two-
dimensional matter is  described by the Lagrangian \cite{Wen2,Moroz}
\begin{eqnarray}
	\label{EFT}
{\cal L}_{{\rm bulk}}=\frac{1}{4\pi} \epsilon^{\mu\nu\rho} a^I_{\mu} K_{IJ}\partial_\nu a^J_{\rho} \nonumber
	- a^I_{\mu} j_{I}^{\mu}\\
	-\frac{1}{2\pi} t_{AI} \epsilon^{\mu\nu\rho}\mathcal{A}^A_{\mu}\partial_\nu a^I_\rho.
		\label{eq:C23}
\end{eqnarray}
Here $a^I$ ($I=1,2,\dots, N$) is a set of gauge fields, $K_{IJ}$ is now an $N\times N$ symmetric integer-number matrix determining the mutual statistics of excitations, $j_I$ are quasiparticle currents, and $t_{AI}$ is a charge vector determining the occupation numbers. The ground state degeneracy  on the torus (which is a characteristic of the topological order) is given by the determinant of the matrix $K$. This determinant also gives the number of independent types of anyons, i.e. fractional charge particles. The last term in (\ref{eq:C23}) describes the coupling to external source fields $\mathcal{A}^A$  ($A=1,2,\dots, M$) which have the global U$(1)_A $ symmetry.

The Hall conductance of the system is quantized if the many-body ground state on a torus is not degenerate and has a finite energy gap. If the many-body ground states have degeneracies, then $K$ and the Hall conductance can be a rational number. If one turns on an interaction, periodic potential, or even a random potential the Hall conductance cannot change.
The only way to change the Hall conductance is to close the energy gap, which means a quantum phase transition.

Fractional quantum Hall states are incompressible and possess a finite energy gap for all their bulk excitations. At the same time, such liquids in finite-size systems always contain one-dimensional gapless edge excitations with a complex structure that reflects a  bulk topological order (the bulk--boundary correspondence). 
In the case of non-abelian fractional quantum Hall liquids, the edge states  produce even more exotic one-dimensional correlated systems \cite{Wen22}. 

Effective field theory (\ref{eq:C23}) provides  understanding of the edge states physics. In the absence of external sources $\mathcal {A}^ A $, a chiral Luttinger theory arises \cite{Moroz,Wen23}: 
\begin{equation}
	\label{edge1}
	{\cal L}_{\rm{edge}}=\frac{1}{4\pi}\Big[K_{IJ}\partial_t \phi^I \partial_x \phi^J-V_{IJ}\partial_x \phi^I \partial_x \phi^J \Big].
\end{equation} 
Here $\phi^I$ are $ N $ chiral bosons, $V_{IJ}$ is a nonuniversal positive definite real matrix depending on the microscopic properties of the edge. 
Thus the above $K$ matrix determinant gives the number of edge states.

In the case of a narrow strongly correlated band, the representation of many-electron Hubbard operators $\tilde{c}_{i \sigma }=X_i(0,\sigma )$ in terms of slave (auxiliary) bosons and fermions can be  used again to describe various classes of quantum Hall spin liquids. This construction has two forms: charged fermionic holons and neutral bosonic (Schwinger) spinons (slave-fermion representation) or neutral fermionic (Abrikosov) spinons and bosonic holons (slave-boson representation). In addition, a U(1) gauge field arises due to the constraint on filling at a site. 
Such a construction (which is also called parton representation) can be introduced for arbitrary symmetry groups describing different types of anyons  \cite{Mao,Ma}. This enables one  to establish the correspondence in the physics of Landau levels and Chern bands, continuous fractional quantum Hall phases, and spin liquids for lattice models.  In this context, phase transitions at half-filling are described as transitions with a change in the Chern number between insulator phases \cite{Parameswaran}.

The topological interaction can be  treated within a quantum field theory (see, e.g., \cite{Oganesyan}).
In 2+1 dimensions,  we have to take into consideration both quasiparticles and vortices, so that  the situation is similar to
the bosonic Chern-Simons  QHE theory and flux and charge can be attributed to the particles in such a way that the Berry phases
occur.  The situation is similar to the  Aharonov-Bohm effect, where   the Fermi--Bose statistics transmutation can take place.
The wavefunction acquires a well-defined Berry phase given by the  integral of the Berry connection. This may be expressed as a surface integral of the Berry curvature, the Chern invariant being the total Berry flux  in the Brillouin zone \cite{Hasan}.

The Chern-Simons Lagrangian \cite{Oganesyan} describes coupling of a  quasiparticle current $j^{\mu}$ and a vortex current $J^{\mu}$   to electric and magnetic gauge potential components $a_{\mu}$ and $b_{\mu}$:
\begin{equation}
	{\cal L} =
	\frac 1 \pi \epsilon^{\mu\nu\sigma}b_\mu\partial_\nu a_\sigma
	- a_{\mu}j^{\mu} - b_{\mu}J^{\mu} \, .
\end{equation}
In 3+1 dimensions, the situation is similar, but we have strings instead of
the vortices,  the vector potential $b$ being  an
antisymmetric tensor $b_{\mu\nu}$.

The slave-fermion representation $\tilde{c}_{i \sigma } = b_{i\sigma} f_i
$ allows one to describe a Z$_2$ spin liquid as singlet pairing of Schwinger bosons with a spin gap and topological order. In the Bose condensate regime, also the antiferromagnetic phase can be described (cf. \cite{Punk}). In this case, the charge response of the Chern insulator is preserved, which just means the state of the quantum Hall spin liquid. The construction of this state is generalized to the fractional filling (fractional QHE). Thus the quantum Hall Z$_2$ spin liquid gives a realization of eight different abelian topological orders with four anyons ($|$det $K|= 4$). In this sense, they are equivalent to eight abelian topological superconductors of  Kitaev's 16-fold way \cite{Kitaev}.

The Fermi spinon representation 
$ \tilde{c}_{i \sigma } = b_i f_{i\sigma}
$
describes a U(1) spin liquid having a spinon Fermi surface. This is the parent state for the Z$_2$ spin liquid which occurs at lowering the gauge symmetry. For even  Chern invariants $C$, the so-called composite Fermi liquid state can be constructed. This is a paramagnetic and metallic analogue of a spin liquid with a purely spinon Fermi surface \cite{Zhang}. Such a state enables one to partially preserve the electronic degrees of freedom, including nonzero residue. 
The formation of this state is owing to the separation of spin and charge degrees of freedom, being connected with strong correlations. 

The bosonic integer QHE (BIQHE) phase presents the simplest  Chern band with an even Chern invariant $C$   \cite{Zhang}. The effective action for this phase reads 
\begin{equation}
	{\cal L}={\cal L}_{BIQHE}[b,A+a]+{\cal L}_{FS}[f,-a].
\end{equation} 
where $A$ is an external U(1) gauge field. One  has in terms of a differential form
\begin{equation}
	\label{biqhe}
	{\cal L}_{BIQHE}= \frac{C}{4\pi}(A+a)d(A+a),
\end{equation}
\begin{equation}
	\label{fs_a}
	{\cal L}_{FS}=f^\dagger_\sigma(\partial_\tau -\mu+ia_0)  f_\sigma-\frac{\hbar^2}{2m^*}f^\dagger_\sigma(-i\partial_i+a_i)^2f_\sigma .
\end{equation} 
Here $\mu$ is the chemical potential, $m^*$ is the effective mass, and $a_0$ and $a_i$ are the temporal and spatial components of the gauge field \cite{Wen2}. The Fermi spinons $ f_\sigma $ partially fill the band without Chern number and form a Fermi surface. 

For odd Chern invariants $C$ = 1, 3, 5, 7, ... , exotic states with the Hall conductivity $\sigma_{xy} = C$ are formed. These are eight types of spin liquids with a half-integer chiral central charge $c = C-1/2$, which are analogous to eight non-abelian Kitaev's superconductors \cite{Kitaev}.

\section{Topological superconductivity} 
\label{sect:supercon}

The experimental discovery of the superconductive order by Kamerlingh Onnes historically resulted in the theory of broken symmetry; however, a later quantum topological investigation complicates this picture. In fact, the superconductivity can be described by the Ginzburg--Landau theory with a dynamic U(1) gauge field \cite{Zoo}. Frequently, the superconductivity is described by the Ginzburg--Landau theory without this field, instead of which a violated symmetry U(1) is treated. 
According to recent theoretical developments  \cite{Wen:1991,Oganesyan}, the real superconductors in an electromagnetic gauge field 
are not states with violated  U(1) symmetry, but can be treated as
topologically ordered states. 
The condensation of an Cooper pair with charge $2e$ breaks the U(1) gauge theory, reducing it to the Z$_2$ gauge theory at low energies.
The latter is the effective theory of topological order Z$_2$, so that a real superconductor has  such a topological order. The superconductor has string excitations, which characterize a topological order -- loops of the flux $hc/2e$.





Topological superconductors have a fully gapped spectrum below the critical  $T_c$, the topological surface states dominating above $T_c$, similar to  topological insulators  \cite{Sato}. 
Topological superconductivity is characteristic for some doped topological insulators.
To realize topological superconductivity, strong spin-orbit coupling, especially for heavy atoms, is favorable. The corresponding examples are some Sn-based superconductors, e.g., Sn$_{1-x}$Sb$_x$  and SnAs \cite{Sharma}.
The presence of essential topology in SnAs is confirmed by first principle calculations of Z$_2$ invariant  \cite{Sharma1}.

 A topologically non-trivial structure exists only in two dimensions (2D), since General lattice Hamiltonians do not have such a structure in one and three-dimensions. Nevertheless, topological superconductors are possible in 1D and 3D, as well as in 2D, owing to the specific symmetry on superconductors \cite{Sato}.
If no symmetry is assumed, the only possible topological phase 
is the quantum Hall state with a non-trivial Chern number.
Thus some additional symmetry is required to obtain other topological phases. 
If the time-reversal $T$ symmetry is present, one can obtain new topological phases, including the quantum spin Hall phase in the 2D case and the topological insulator phase in the 3D case.

For the 2D case, the particle-hole symmetry does not result in a new topological structure.
Similar to the quantum Hall or spin Hall states, one can define the Chern number or the 2D Z$_2$ index (-1)$^{\nu_{2D}}$ in the absence or presence of $T$ symmetry, respectively. 

In the presence of $T$ symmetry, the 3D Z$_2$ index  
can be introduced in the same way as in topological insulators, but the combination with particle-hole symmetry enables one to introduce a more sophisticated integer topological number \cite{Sato}. 
Thus the particle-hole symmetry plays an important role in the 3D case.  

Due to the Higgs phenomenon, 
the ground state spectrum of a two-dimensional superconductors with electrically charged paired fermions interacting with a dynamical  electromagnetic field has a gap. 
This means a description in terms of 2D Z$_2$ spin liquid within the framework of effective Z$_2$ gauge theory with a Z$_2$ topological
order. According to \cite{Oganesyan,Zoo}, this construction can be  generalized
to treat 3D Z$_2$ spin liquid, its Z$_2$ topological order being the same as in an $s$-wave superconductor.


A 2D superconductor possesses two types of point excitations: Bogoliubov quasiparticles and vortices, their braiding being possible. 
Moroz et al \cite{Moroz} developed a low-energy description for spin-singlet pairing states by elaborating  Chern-Simons field theories for even-wave superconductors: $s$-wave, $d+id$, etc.  Combining topological order and symmetries in superconductors, this approach
demonstrates that these systems are symmetry-enriched topological (SET) phases discussed in the Introduction.

The corresponding theories of chiral spin-singlet superconductors are described by Kitaev's  classification of topological superconductors (the ``16-fold way'').
For chiral spin-singlet superconductors having the chirality parameter $k$ (i.e.,   
the orbital angular momentum of a Cooper pair, which is an even integer (0 for $s$-pairing and 2 for $d$-pairing), one finds $c = k$.
According to Kitaev \cite{Kitaev}, chiral superconductors have a Z$_{16}$ bulk classification. Since each Majorana mode gives a half unit of the central charge, the total chiral central charge is $c = \nu/2$ with $\nu$ the Chern number.

For $s$- and $d-$wave superconductor one has the four-component theory with the resulting $K$-matrix, respectively 
\begin{equation} \label{Ks}
K=\left(
\begin{tabular}{ ccc c}
0 & 2 &1 & 1 \\
2 & 0 &0 & 0  \\
1 & 0 &1 & 0 \\
 1 & 0 &0 & $\mp 1$ \\
\end{tabular} 
\right).
\end{equation}
This $K$-matrix  contains the  bosonic block $\big(
\begin{tabular}{ cc}
	0 & 2 \\
	2 & 0  \\
\end{tabular}
\big).$ 
In the case of a $s$-wave superconductor,  Bogoliubov particles carry  spin and vortices  -- magnetic flux,  their composite being a boson which bears both  spin and vortex charge. Their mutual braiding gives the statistical angle $\pi$:  
the Bogoliubov quasiparticle  accumulates a minus sign when encircling a vortex.

In a chiral $d + id$ superconductor, parity and $T$ symmetry are violated spontaneously leading thereby to
anyon statistics of excitations.
Now the vortex  is a semion, which is demonstrated by the Berry
phase gained via exchange of two identical vortices. 

For a spin-singlet  superconductor, the chiral central charge
$c$ is equal to the chirality parameter $k$.
The states with  an odd  $c$ constitute another class of abelian
states  probably describing some  spin-triplet superconducting phases \cite{Moroz}.
Thus spin-triplet superconducting pairing is important from the topological point of view, since it can mean existence of topological states and Majorana fermions.
Recently, superconducting compound UTe$_2$  has been supposed to be a candidate for a chiral spin-triplet topological superconductor near a ferromagnetic instability  \cite{Podlesnyak}.  Inelastic neutron scattering  demonstrates that superconductivity  in UTe$_2$  near antiferromagnetic order is coupled to a sharp magnetic excitation (a resonance) at the Brillouin zone boundary. This suggests that antiferromagnetic spin fluctuations may  lead to spin-triplet pairing.

A  doping-induced  transition from Mott insulator to superconductor occurs naturally provided that the parent Mott insulator is a Z$_2$  spin liquid. On the other hand, doping a U(1) spin liquid  naturally provides a Fermi liquid phase,  and not a superconductor, since there is no pairing of spinons in the parent Mott insulator \cite{Senthil2,Song}.

According ot Ref. \cite{Song}, in the case of the triangular lattice
 two different chiral spin liquid phases are possible:  CSL1 which
 is an analogue of fractional quantum Hall state in spin
 system, and CSL2 is a ``gauged'' chiral superconductor, the $\nu=4$ member of the Kitaev  classification.
 Both CSL1 and CSL2 are described in mean-field approach of Abrikosov  fermions, and the doping is treated by introducing charged holons.
 In CSL1, the Fermi spinons are in a Chern insulator state with $C = 2$.  For CSL2, the spinon is in a  $d + id$ superconductor state, which is  similar  to a Z$_2$ spin liquid. Indeed, a Z$_2$  spin liquid can be treated in terms of Fermi spinons in a BCS state. Doped charges are treated in terms of Bose holons, which can condense. Holon condensation transforms the spinon pairing    into usual Cooper electron one, which results in a superconducting phase. 

At doping CSL2,   holons are condensed to form a topological $d+id$ superconductor. At doping CSL1,  two scenarios are possible. Holon condensation results in a chiral metal with doubled unit cell and finite Hall conductivity. In the second  scenario, the internal magnetic flux is set with doping and holons form a bosonic integer quantum Hall (BIQH) state which is identical to a $d+id$ superconductor.


\section{Conclusions}
\label{sec:2}

In the present paper, we have considered various aspects of quantum topological states. This issue should be distinguished from the problems of the Fermi surface topology  \cite{Volovik}: their phenomenological treatment,  as well as physics of topological insulators \cite{Hasan}, is close to classical topology.

The traditional approach to the problem of phase transitions was based on the Landau broken-symmetry theory. 
The new theoretical developments in the condensed matter physics are connected
with new kinds of orders and classes of matter which can violate the Landau theory. Moreover, there exist  various combinations of symmetry breaking and topological orders (long-range entanglements) which occur together in such states as SB-LRE (see the Introduction). The topological superconducting states provide examples of such phases \cite{Zoo}.

The occurrence of exotic states is possible too. For example, in Ref. \cite{kagome}, a Chern-Simons theory was constructed for the doped spin-1/2 kagome system.  This  system turns put to be an exotic superconductor that violates time-reversal symmetry,  carries minimal vortices of flux $hc/4e$ and, in addition to the usual spin-1/2 fermionic Bogoliubov quasiparticles, contains fractional excitations. These are Fermi quasiparticles with semion mutual statistics and spin-1/2 quasiparticles with Bose self-statistics.

It seems that there is a whole new world before us, which is to be
investigated. The new topological paradigm may  contribute to our
understanding of fundamental questions of nature (see discussion in Refs. \cite{Wen22,Zeng,Scr2}).
Most interesting results have been up to now obtained for 2D quantum systems, whereas a detailed classification of 3D ones is an open problem \cite{Zoo}.

The research was carried out within the state assignment of the Ministry of Science and Higher Education of the Russian Federation (theme ``Flux'' No 122021000031-8 and theme ``Quantum'' No. 122021000038-7). 



%
%


\begin{thebibliography}{}
	
\bibitem{Tsui}
D. C. Tsui, H. L. Stormer, and A. C. Gossard, 
Phys. Rev. Lett. \textbf{48}, 1559–1562 (1982).


\bibitem{Bednorz}
J. G. Bednorz and K. A. Mueller, 
Z. Phys. B \textbf{64}, 189 (1986).
%

\bibitem{Kalmeyer}
V. Kalmeyer and R. B. Laughlin,
Phys. Rev. Lett. \textbf{59}, 2095 (1987).

\bibitem{Witten}
E. Witten, 
Commun. Math. Phys. \textbf{121}, 351 (1989).

\bibitem{Zoo}
X. G. Wen, 
Rev. Mod. Phys. \textbf{89}, 41004 (2017).

\bibitem{Mermin}
N. D. Mermin, 
Rev. Mod. Phys. \textbf{51}, 591 (1979).

\bibitem{Zeng}
B. Zeng, X. Chen, D-L. Zhou, and X-G. Wen, {\it Quantum
	information meets quantum matter, from quantum
	entanglement to topological phase in many-body systems,
	In the Springer Book Series—Quantum Information
	Science and Technology};
arXiv:1508.02595 (2015).

\bibitem{1406.5090}
B. Zeng and X. G. Wen, 
Phys. Rev. B \textbf{91}, 125121  (2015).

\bibitem{cre}
X. Chen, Zh.-Ch. Gu,  and X.-G. Wen, Phys. Rev. B \textbf{82}, 155138 (2010).

\bibitem{cre1}
X. Chen, Zh.-Ch. Gu, Zh.-X. Liu, and X.-G. Wen, Phys. Rev. B \textbf{87}, 155114 (2013).

\bibitem{Scr2}
V. Yu. Irkhin and Yu. N. Skryabin, Phys. Met. Metallogr. \textbf{120}, 513 (2019).

\bibitem{Scr3}
V. Yu. Irkhin and Yu. N. Skryabin, J. Exp. Theor. Phys.\textbf{133},  116 (2021).

\bibitem{Lieb}
E. H. Lieb, T. D. Schultz, and D. C. Mattis, Ann. Phys.
(N.Y.) \textbf{16}, 407 (1961).

\bibitem{Hastings}
M. B. Hastings, Phys. Rev. B \textbf{69}, 104431 (2004)

\bibitem{wen2002quantum}
X.-G. Wen, Phys. Rev. B, \textbf{65}, 165113 (2002).

\bibitem{Wen22}
X.-G. Wen, {\it Quantum Field Theory of Many-body Systems}, Oxford University Press, Oxford, 2004.


\bibitem{Nagaosa}
P.A. Lee, N. Nagaosa, and X.-G. Wen, Rev. Mod. Phys. \textbf{78}, 17  (2006).

\bibitem{Weng}
P. Ye, C.-Sh. Tian, X.-L. Qi, and Zh. Weng, 
Nucl. Phys. B\textbf{854}, 815 (2012).

\bibitem{Sachdev}
T. Senthil, M. Vojta, S. Sachdev, Phys. Rev. B \textbf{69}, 035111 (2004).

\bibitem{Punk1}
S. Sachdev, M. A. Metlitski, M. Punk, J. Phys.: Cond. Mat. \textbf{24}, 294205 (2012).

\bibitem{633a} P. W.  Anderson, Science \textbf{235} 1196 (1987).

\bibitem{Baskaran}
G. Baskaran, Z. Zou, and P. W. Anderson, Solid State
Comm. \textbf{63}, 973 (1987).


\bibitem{Sachdev2}
S.  Sachdev, In: "Quantum magnetism", U. Schollwock, J. Richter, D. J. J. Farnell and R. A. Bishop eds, Lecture Notes in Physics, Springer, Berlin (2004); arXiv:cond-mat/0401041.

\bibitem{Wen02}
X.-G. Wen, Phys. Rev. B\textbf{65}, 165113 (2002).

\bibitem{Sachdev12} S. Sachdev, arXiv:1203.4565.

\bibitem{Scr4}
V. Yu. Irkhin and Yu. N. Skryabin, Phys. Lett. A \textbf{383}, 2974  (2019).

\bibitem{Hubbard}
J. Hubbard,
Proc. Roy. Soc. A \textbf{276}, 238 (1963). 

\bibitem{Hubbard2}
J. Hubbard, Proc. Roy. Soc. A \textbf{277}, 237 (1963).


\bibitem{Hubbard3}
J. Hubbard,
Proc. Roy. Soc. A \textbf{281}, 401 (1964).

\bibitem{Levin}
M. A. Levin and X. G. Wen, 
Phys. Rev. B \textbf{71}, 045110 (2005).

\bibitem{Levin1}
Ch. Chen, P. Rao, I. Sodemann, 	arXiv:2202.01238.

\bibitem{Weng1} 
W. Zheng, Zh. Zhu, D. N. Sheng, and Zh.-Y. Weng, Phys. Rev. B \textbf{98}, 165102 (2018).

\bibitem{Weng2007}
Z.-Y. Weng,
Int. J.  Mod. Phys. B \textbf{21},
773 (2007).

\bibitem{Zaanen-Overbosch}
J. Zaanen and B. J. Overbosch,
Phil. Trans. R. Soc. A  \textbf{369}, 1599 (2011).


\bibitem{Sachdev1}
T. Senthil, M. Vojta, S. Sachdev, Physica B \textbf{359-361}, 9 (2005).

\bibitem{Vojta} M. Vojta, Rep. Prog. Phys. \textbf{81}, 064501 (2018).


\bibitem{Oshikawa}
M. Oshikawa, Phys. Rev. Lett. \textbf{84}, 3370 (2000).

\bibitem{Senthil1}
T. Senthil,
Phys. Rev. B \textbf{78}, 035103 (2008).

\bibitem{Senthil2}
T. Senthil,
Phys. Rev. B \textbf{78}, 045109 (2008).

\bibitem{Kogut}
J. B. Kogut, 
Rev. Mod. Phys. \textbf{51}, 659 (1979).

\bibitem{Kitaev}
A. Kitaev, Ann. Phys. (N. Y.) \textbf{321}, 2 (2006).

\bibitem{Hasan} M. Z. Hasan and C. L. Kane, Rev. Mod. Phys. \textbf{82}, 3045 (2010).

\bibitem{Haldane}
F. D. M. Haldane, Phys. Rev. Lett. \textbf{61}, 2015 (1988).

\bibitem{Neupert}
T. Neupert, L. Santos, C. Chamon, and C. Mudry,
Phys. Rev. Lett. \textbf{106}, 236804 (2011).

\bibitem{Tang}	
E. Tang, J.-W. Mei, and X.-G. Wen. Phys. Rev. Lett. \textbf{106} 236802 (2011). 

\bibitem{Zhu}
W. Zhu, S. S. Gong, and D. N. Sheng,
Phys. Rev. B \textbf{94}, 035129 (2016).

\bibitem{Zhang}
Y.-H. Zhang and T. Senthil,
Phys. Rev. B \textbf{102}, 115127 (2020).



\bibitem{Wen2}
X.-G. Wen, Adv. Phys. \textbf{44}, 405 (1995).

\bibitem{Oganesyan} T. H. Hansson, V. Oganesyan, S. L. Sondhi, Ann. Phys. \textbf{313}, 497 (2004).


\bibitem{Moroz}
S. Moroz, A. Prem, V. Gurarie, L. Radzihovsky, Phys. Rev. B \textbf{95}, 014508 (2017).


\bibitem{Wen23}
X.-G. Wen, Int. J. Mod. Phys. B\textbf{6}, 1711 (1992).

\bibitem{Mao}
Y.-H. Zhang, D. Mao, Phys. Rev. B \textbf{101}, 035122 (2020).

\bibitem{Ma}
R. Ma, Y.-Ch. He,  Phys. Rev. Research \textbf{2}, 033348 (2020).

\bibitem{Parameswaran}
S. A. Parameswaran, R. Roy, Sh.L. Sondhi, C. R. Physique \textbf{14}, 816 (2013).

\bibitem{Punk} M. Punk and S. Sachdev, Phys. Rev. B \textbf{85}, 195123 (2012).



\bibitem{Wen:1991}
X.G. Wen,
Phys. Rev. B \textbf{44}, 2664 (1991).


\bibitem{Sato}
M. Sato and Y. Ando, Rep. Prog. Phys. \textbf{80}, 076501 (2017). 

\bibitem{Sharma}
M. M. Sharma, V.P.S. Awana,	arXiv:2202.10664.

\bibitem{Sharma1}
M. M. Sharma, N. K. Karn, P. Sharma, G. Gurjar, S. Patnaik, V.P.S. Awana, Solid State Commun. \textbf{340}, 114531 (2021).

\bibitem{Podlesnyak}
Ch. Duan, R. E. Baumbach, A. Podlesnyak, Y. Deng, C. Moir, A. J. Breindel, M. Brian Maple, E. M. Nica, Q. Si, and P. Dai,
Nature \textbf{600}, 636 (2021).

\bibitem{Song}
X.-Y. Song, A. Vishwanath, and Y.-H. Zhang,
Phys. Rev. B \textbf{103}, 165138 (2021).

\bibitem{Volovik}
G. E. Volovik, Phys.-Usp. \textbf{61}, 89 (2018).

\bibitem{kagome}
W.-H. Ko, P. A. Lee, and X.-G. Wen,
Phys. Rev. B \textbf{79}, 214502 (2009).

\end{thebibliography}


\end{document}